\documentclass[aps,pra,epsfig,graphics,showpacs,twocolumn,floatfix,mathbbm,a4paper]{revtex4}

\usepackage{amsmath,amsfonts,amssymb,graphics,graphicx,epsfig,color,times,bbm}

\begin{document}
\bibliographystyle{apsrev}

\newcommand{\R}{\mathbbm{R}}
\newcommand{\rr}{\mathbbm{R}}
\newcommand{\nn}{\mathbbm{N}}
\newcommand{\cc}{\mathbbm{C}}
\newcommand{\ii}{\mathbbm{1}}

\newcommand{\id}{\mathbbm{1}}

\newcommand{\tr}{{\rm tr}}
\newcommand{\gr}[1]{\boldsymbol{#1}}
\newcommand{\be}{\begin{equation}}
\newcommand{\ee}{\end{equation}}
\newcommand{\bea}{\begin{eqnarray}}
\newcommand{\eea}{\end{eqnarray}}
\newcommand{\ket}[1]{|#1\rangle}
\newcommand{\bra}[1]{\langle#1|}
\newcommand{\avr}[1]{\langle#1\rangle}
\newcommand{\G}{{\cal G}}
\newcommand{\eq}[1]{Eq.~(\ref{#1})}
\newcommand{\ineq}[1]{Ineq.~(\ref{#1})}
\newcommand{\sirsection}[1]{\section{\large \sf \textbf{#1}}}
\newcommand{\sirsubsection}[1]{\subsection{\normalsize \sf \textbf{#1}}}
\newcommand{\ack}{\subsection*{\normalsize \sf \textbf{Acknowledgements}}}
\newcommand{\front}[5]{\title{\sf \textbf{\Large #1}}
\author{#2 \vspace*{.4cm}\\
\footnotesize #3}
\date{\footnotesize \sf \begin{quote}
\hspace*{.2cm}#4 \end{quote}
#5} \maketitle}
\newcommand{\eg}{\emph{e.g.}~}

\newcommand{\proofend}{\hfill\fbox\\\medskip }

%---------------------------------------------------------------------------

\newtheorem{theorem}{Theorem}
\newtheorem{proposition}{Proposition}

\newtheorem{lemma}{Proposition}

\newtheorem{definition}{Definition}
\newtheorem{corollary}{Corollary}

\newcommand{\proof}[1]{{\bf Proof.} #1 $\proofend$}

\title{Multiplicativity of maximal output purities\\
of Gaussian channels under Gaussian inputs}

\author{A.~Serafini$^{1,2}$, J.~Eisert$^{2,3}$ and M.M.~Wolf$^{4}$}

\address{$^1$ Dipartimento di Fisica ``E.R.~Caianiello'', Universit\`a di
Salerno,
INFM UdR Salerno, INFN Sezione di Napoli, Gruppo Collegato di Salerno, Via
S.~Allende,
84081, Baronissi (SA), Italy}

\address{$^2$ Institut f{\"u}r Physik, Universit{\"a}t Potsdam,
Am Neuen Palais 10, D-14469 Potsdam, Germany}

\address{$^3$ Blackett Laboratory, Imperial College London,
Prince Consort Road, London SW7 2BW, UK}
\address{$^4$ Max-Planck-Institut f\"ur Quantenoptik, Hans-Kopfermann-Stra{\ss}e
1, 85748 Garching, Germany}

\date{\today}

%\maketitle

\begin{abstract}
We address the question of the multiplicativity of the maximal
$p$-norm output purities
 of bosonic
Gaussian channels under Gaussian inputs. We focus on general
Gaussian channels resulting from the reduction of unitary dynamics
in larger Hilbert spaces. It is shown that the maximal output
purity of tensor products of
 single-mode channels under Gaussian inputs is
multiplicative for any $p\in(1,\infty)$ for products of arbitrary
identical channels as well as for a large class of products of
different channels. In the case of $p=2$  multiplicativity is
shown to be true for arbitrary products of generic channels acting
on any number of modes.
%Our results suggest the general
%multiplicativity of the Gaussian maximal output purity for any $p$
%and tensor product of single-mode channels.
\end{abstract}


\maketitle

%%%%%%%%%%%%%%%%%%%%%%%%%%%%%%%%%%%%%%%%%%%%%%%%%%%%%%%%%%%%%%%%%%%%%%

%\begin{multicols}{2}

\section{Introduction}

Additivity and multiplicativity questions play a central role in
the field of quantum information theory: does it help to make
joint use of a quantum channel when transmitting quantum or
classical information in form of entangled inputs, or is one
better off by merely invoking the channel many times with
uncorrelated inputs? Or, what is the entanglement cost, the rate
at which maximally entangled pairs need to be invested in the
asymptotic preparation of a mixed bi-partite state using local
operations and classical communication only? If the so-called
entanglement of formation turned out to be additive, the
evaluation of this quantity, which amounts to a much simpler
optimization problem, would be sufficient to provide the complete
answer to this question. Most instances of such additivity
problems in quantum information share the common feature of being
notoriously difficult to solve. Recently, yet, a picture emerged
that made clear that several of these problems share more than a
formal similarity \cite{Shor,conse,amosov,Multiplicative}.
It has actually been shown that at least four instances of such
additivity problems are logically equivalent, being either all
wrong or all true. Besides the fundamental insight that this equivalence
provides, such an observation is practically helpful, since it links
isolated additivity results to other instances of such problems.

This paper is concerned with a specific multiplicativity question
for quantum channels of bosonic systems: it deals with the maximal
output purity of Gaussian channels under Gaussian inputs. This
quantity specifies how well the purity of an input state, measured
in terms of $p$-norm purities (or equivalently R\'enyi entropies),
can be preserved under the application of the (generally
decohering) channel and provides a way to characterize the
decoherence rate of the channel. If this output purity turns out
to be multiplicative for a tensor product of channels, then input
entanglement cannot help to better preserve the coherence of the
output states \cite{GeneralRemark}. The multiplicativity of the
maximal output purity for $p\rightarrow1+$ corresponds to the
additivity of the minimal von Neumann entropy. Furthermore, if
general inputs are allowed for, the multiplicativity for such a
limiting instance is strictly related to the additivity of the
(appropriately constrained) Holevo capacity and of the
entanglement of formation (EoF) \cite{Shor,conse,amosov}.

In turn, the quantum information properties of bosonic Gaussian
channels \cite{Channels,capa} have attracted strong theoretical
attention in recent years. This class of quantum channels is
practically very important: indeed, the transmission of light
through a fiber is described to a very good approximation by a
Gaussian bosonic channel. The unavoidable coupling to external
field modes yield losses, whereas excess noise can be incorporated
as random classical Gaussian noise, reflecting random
displacements in phase space. The estimation of various
information capacities, both quantum and classical, has been
thoroughly addressed for these Gaussian channels \cite{capa}. As
for the multiplicativity of the maximal output purity, the quest
has been challenged in a series of recent works, and strong
arguments have been provided to support it, addressing a subset of
channels investigated in the present paper \cite{lloyd}. In fact,
for a specific channel model describing a beam splitter
interaction of a bosonic mode with a thermal noise source and for
integer $p\geq 2$ multiplicativity of the maximal $p$-norm output
purity was recently proven \cite{lloyd}. However, a definitive
proof of the multiplicativity conjecture for tensor products of
 general Gaussian channels and non-integer $p$ is still missing.

In the present paper we aim to make a step towards a theory of
channel capacities of Gaussian channels under Gaussian inputs,
dealing with more general instances of channels. Such a setting,
besides being interesting in its own right, yields obvious bounds
for the unconstrained maximal output purity for Gaussian bosonic
channels, which in turn may be conjectured to be tight as it is
the case for particular single-mode channels and integer $p\geq 2$
\cite{lloyd}. Moreover, the present paper is meant to be a further
step towards a clear picture of a general quantum information
theory of Gaussian states, linking channel capacities with
entanglement properties. This picture could provide a powerful
laboratory, when a complete solution to the specified additivity
problems is lacking.

The paper is structured as follows.
In section \ref{gauss} we introduce the notation, basic facts about
Gaussian states and
we define the class of Gaussian channels we will deal with. In section
\ref{purity}
the $p$-norms as measures of purity are presented and determined for
Gaussian states.
In section \ref{mop} the Gaussian multiplicativity of the maximal output
purity is
defined, while section \ref{theo} contains all the analytical results
about
multiplicativity. Finally, in section
\ref{comme} we review our results and provide some comments and
perspectives.

\section{Gaussian states and channels \label{gauss}}

\subsection{Gaussian states}

We consider quantum systems with $n$ canonical degrees of freedom, i.e.,
a system consisting of $n$ modes.
The canonical coordinates corresponding to position and momentum will
be denoted as
$\hat R=(\hat x_1, \hat p_1,...,\hat x_n,\hat p_n)$,
where in terms of
the usual creation and annihilation operators we have that
$\hat x_i =( \hat a_i + \hat a_i^\dagger)/\sqrt{2}$ and
$\hat p_i=-i(\hat a_i - \hat a_i^\dagger)/\sqrt{2}$.
 In terms of the
Weyl operators
\begin{equation}
    W_\xi = e^{ i \xi^T \sigma \hat R}, \,\,\, \xi \in \rr^{2n}
\end{equation}
the canonical commutation relations (CCR) can be written as
\begin{equation}
    W_\xi^\dagger W_{\xi'} =W_\xi^\dagger W_{\xi'}
     e^{i \xi^T \sigma \xi'},
\end{equation}
where
\begin{equation}\label{symform}
    \sigma = \bigoplus_{i=1}^n
    \left(
    \begin{array}{cc}
    0 & 1\\
    -1 & 0 \\
    \end{array}
    \right).
\end{equation}
The latter matrix $\sigma$ is the symplectic matrix.
States can be fully characterized by functions in
phase space $(\rr^{2n},\sigma)$. The characteristic function is
defined as
\begin{equation}
    \chi_\rho (\xi) = {\rm tr}[ \rho W_\xi ],
\end{equation}
where the state $\rho$ can in turn be expressed as
\begin{equation}
    \rho=\frac{1}{(2\pi)^n}\int
    d^{2n} \xi
    \chi_\rho(-\xi) W_\xi;
\end{equation}
The characteristic function is the ordinary Fourier transform of
the Wigner function commonly employed in the phase space
description of quantum optics \cite{GZ, book}.

Gaussian states are, by definition, the states with Gaussian
characteristic function (and therefore Gaussian Wigner function)
\begin{equation}
\chi(\xi)=\chi(0){\rm e}^{-\frac12\xi\Gamma\xi^T+\xi D}\; .
\end{equation}
Gaussian states are fully determined by first and second moments
of the quadrature operators, respectively embodied by the vector
$d=\sigma D$ and by the real symmetric $2n\times 2n$ matrix
$\Gamma=\sigma^T \gamma\sigma$, with
\begin{equation}
\gamma_{ij}=\frac12\langle\hat{R}_i\hat{R}_j+\hat{R}_j\hat{R}_i\rangle_{\rho}
-\langle\hat{R}_i\rangle_{\rho}\langle\hat{R}_j\rangle_{\rho}
\, ,
\end{equation}
where $\langle\hat{O}\rangle_{\rho}=\,{\rm tr}\,[\rho
\hat{O}]$ for the operator $\hat{O}$. First moments can be set to zero
by local unitaries, so that they play no direct role in properties
related to entanglement and mixedness of Gaussian states. To our
aims a Gaussian state will be characterized by its covariance
matrix  $\gamma$.
%actually representing
%the class of Gaussian states with such second moments.
The covariance matrix $\gamma$ of a Gaussian state $\rho$ (and, indeed, any
covariance matrix related to a physical state), has to satisfy
the uncertainty principle
\begin{equation}
\gamma + i \sigma\geq0 \; , \label{heis}
\end{equation}
reflecting the positivity of $\rho$.
Subsequently, ${\cal G}$ will stand for the set of Gaussian states
with vanishing first moments  (for simplicity,
the underlying number of canonical degrees of freedom
will not be made explicit).
Pure Gaussian states
are those for which $\det\gamma=1$. These are the minimal uncertainty states,
saturating Ineq.~(\ref{heis}).
The subset of pure Gaussian states with vanishing first moments will be
denoted as $\tilde{\G}$.

Any unitary  $U$  generated by
polynomials of degree two
in the canonical coordinates is, by virtue of the Stone--von Neumann
theorem, the metaplectical representation
of a real symplectic transformation $S\in Sp(2n,{\mathbbm R})$.
We recall that the real symplectic
group $Sp(2n,\rr)$
consist of those real $2n\times 2n$ matrices
$S$ for which $S^T\sigma S=\sigma$.
Such symplectic operations preserve the Gaussian character of the
states and act by congruence on covariance matrices
\begin{equation}
\gamma \longmapsto S^T \gamma S \; .
\end{equation}
On Weyl operators such an operation is
reflected by
\begin{equation}
    W_\xi \longmapsto W_{S^{-1} \xi}\; .
\end{equation}
We mention that
ideal beam splitters and squeezers are described by symplectic transformations.
The expression of the generators of $Sp(2n,\R)$ will be useful in the
following and is detailed in App.\ \ref{genius}.
Moreover, we recall that  a useful
way to express a generic symplectic transformation $S$ is provided by the
Euler decomposition \cite{prama}
\begin{equation}
S = O' Z O'' ,\label{euler1}
\end{equation}
where $O', O'' \in K(n)= Sp(2n,\R)\cap SO(2n)$ are orthogonal
symplectic transformations, whose set forms the maximal compact subgroup
of $Sp(2n,\R)$. They are those operations typically referred to as being
passive, again, in optical systems corresponding to
beam splitters and phase shifts. The group of all
$Z=\,{\rm diag}\,(z_1,1/z_1,\ldots,z_n,1/z_n)$
with $z_1,...,z_n\in \R^{+} \backslash \{0\}$ is
the non-compact group of all such $Z$, reflecting local squeezings;
this group will be denoted by $Z(n)$ in the following.

A frequently
used tool will be the fact that any CM $\gamma$ can be brought
to the Williamson normal form \cite{williamson36}
\begin{equation}\label{willia}
    \gamma \longmapsto S\gamma S^T = \bigoplus_{i=1}^n
    \left(
    \begin{array}{cc}
    \nu_i^\downarrow & 0 \\
    0 & \nu_i^\downarrow
    \end{array}
    \right),
\end{equation}
with $\nu_i^\downarrow \in [1,\infty)$ and $S\in Sp(2n,\rr)$. The
vector $(\nu^\downarrow_1,...,\nu^\downarrow_n)$ is the vector of
decreasingly ordered symplectic eigenvalues, which can be computed as
the spectrum of the matrix $|i \sigma \gamma|$.
The previous decomposition is nothing but
the normal mode decomposition.
Choosing the standard number basis $\{ |n\rangle :n\in \nn \}$
of the Hilbert space associated with each mode, the
Gaussian state with vanishing first moments and
the second moments as in the right hand side of
Eq.\ (\ref{willia}) is given by
\begin{equation}
\rho=\bigotimes_{i=1}^n
\frac{2}{\nu_{i}^\downarrow+1}\sum_{k=0}^{\infty}\left(
\frac{\nu_{k}^\downarrow-1}{\nu_{k}^\downarrow+1}\right)^k\ket{k}
\bra{k}\; . \label{thermas}
\end{equation}

Recalling the Euler decomposition and Williamson's theorem, expressed by
\eq{euler1} and \eq{willia} one then finds that the CM $\gamma$
of an arbitrary pure Gaussian state  reads
\begin{eqnarray}
\gamma = O^T Z O \; , {\rm with}\quad O\in K(n) \, , \; Z \in Z(n)\, .
\label{euler2}
\end{eqnarray}
We finally mention that, as it is evident from the definition of the
characteristic function,
tensor products of Hilbert spaces correspond to direct sums of phase
spaces. Therefore
an uncorrelated tensor product of Gaussian states
with CMs $\gamma_i$, $i=1,...,n$, has the
CM $\gamma=\oplus_{i=1}^n
\gamma_i$. Likewise, for a `local' tensor product of symplectic
transformations $S_i$ one has
$S=\oplus_{i=1}^n S_i$.

\subsection{Gaussian channels}

In general, a Gaussian channel is a trace-preserving completely positive map that
maps Gaussian trace-class operators onto Gaussian trace-class operators.
A Gaussian channel  is defined by its action on the Weyl operators, according to
\begin{equation}
    W_\xi \longmapsto
    W_{X\xi} e^{- y(\xi)},
\end{equation}
where $X$ is a real $2n\times2n$-matrix, and $y$ is a quadratic
form. We do not consider linear terms in this quadratic form,
which would merely correspond to a displacement, i.e., a change in
first moments. We can hence write $y(\xi)=  \xi^T Y \xi/2$.
Complete positivity of the channel requires that
\begin{equation}
    Y + i \sigma - i X^T \sigma X\geq 0 \; .
\end{equation}
For single-mode channels, this requirement is equivalent to
\begin{equation}
    Y\geq 0,\,\,\,
    \det [Y] \geq (\det [X] -1)^2\; . \label{pos2cha}
\end{equation}
The second moments are transformed under the application of such a
channel according to
\begin{equation}
    \gamma\longmapsto X^T\gamma X + Y\; . \label{channel}
\end{equation}
%This is a quite general class of Gaussian memoryless channels
%(with the sole exception of operations involving projections on
%singular `position eigenstates', giving rise to a more general class
%of Gaussian trace preserving maps, see Ref.~\cite{book}).
Any channel of the form of \eq{channel} corresponds to the
reduction of a symplectic (unitary) evolution acting on a larger
Hilbert space and, vice versa, any evolution of this kind is
described by \eq{channel} for some $X$ and $Y$ \cite{book}.

In the Schr{\"o}dinger picture, we will denote such channels
(characterized by the matrices $X$ and $Y$) by $\Phi_{X,Y}$,
acting as
\begin{equation}
    \rho\longmapsto \Phi_{X,Y}(\rho)\; .
\end{equation}
This class of channels includes the
classical case of random displacements with a Gaussian weight
\begin{equation}\label{averagedgamma}
\rho \longmapsto \int d^{2n}\xi P(\xi) W_{\xi}^{\dag}\rho W_{\xi}
 \, ,
\end{equation}
where $P(\xi)=P(0)\,{\rm e}^{-\frac12 \xi Y^{-1}\xi^T}$ is a
multivariate Gaussian with positive covariance matrix $Y$. In our
notation, such a channel corresponds to
$
X=\ii,\,\,\,Y\geq 0.
$
The amplification and attentuation channels can be described by \eq{channel}
too, with
\begin{equation}
    X= \varepsilon \ii_2 ,\,\,\,Y=|1-\varepsilon^2|\ii_2 \, ,
\end{equation}
with $\varepsilon<1$ (attenuation) or $\varepsilon>1$ (amplification).
Note that this instance encompasses the case of white noise as well.

Also the dissipation in Gaussian reservoirs after a time $t$ is
included in such a class of channels, with the choices \be
X=\,{\rm e}^{-\Gamma t/2}{\mathbbm 1}\, , \; Y=(1-\,{\rm
e}^{-\Gamma t})\gamma_B\, ,\label{bath} \ee where $\Gamma$ is the
coupling to the bath and $\gamma_B$ is the covariance matrix
describing the reservoir.

The channel model on which the papers of Ref.~\cite{lloyd} are
focused on is characterized by $X=c\ii, c\leq 1$ and $Y$ diagonal (mainly $Y\propto \ii$).\\

The additional noise term $Y$ of a general channel incorporates
both the noise that is due to the Heisenberg uncertainty in a
dilation, and the additional classical noise. In the same manner
as minimal uncertainty Gaussian states can be introduced, pure
channels can be considered, satisfying
\begin{equation}
    Y=- (X^T \sigma X-\sigma) Y^{-1} (X^T \sigma X-\sigma),
\end{equation}
where the inverse has to be understood as the Moore-Penrose inverse.

\section{Measures of purity \label{purity}}

Generally, the degree of purity of a quantum state
$\rho$ can be characterized by its Schatten $p$--norm
\cite{bathia}
\be
\|\rho\|_p=(\,{\rm
tr}\,|\rho|^p)^{\frac1p} =(\,{\rm tr}\,\rho^p)^{\frac1p}\, ,
\quad \, p\in (1,\infty).
 \ee
We mention that the case $p=2$ is directly related to the quantity
often referred to as linear entropy or purity in the closer sense,
$\mu=\,{\rm tr}\,\rho^2=\|\rho\|_{2}^{2}$. The $p$-norms are
multiplicative on tensor product states and determine the family
of R\'enyi entropies $S_{p}$ \cite{renyi}, given by \be
S_{p}=\frac{\ln\,{\rm tr}\,\rho^p}{1-p} \; , \label{pgen} \ee
quantifying the degree of mixedness of the state $\rho$. It can
be easily shown that \be \lim_{p\rightarrow1+}S_{p}=-\,{\rm
tr}\,[\rho\ln\rho] = S_{V}(\rho) \, . \label{genvneu} \ee
Thus the von-Neumann entropy $S_{V}$ is determined by $p$-norms,
as it is given by the first derivative of $\| \rho\|_p$ at
$p\rightarrow1+$. The quantities $S_p$ are additive on tensor
product states. It is easily seen that $S_p(\rho) \in
(0,\infty)$,  taking the value $0$ exactly on pure states.

Because of the unitary invariance of the $p$-norms,
the quantities
${\rm tr}\,\rho^p$ of a $n$-mode Gaussian
state $\rho$
can be simply computed in terms of its symplectic
eigenvalues.
In fact, due to \eq{willia},    ${\rm
tr}\,\rho^{p}$ can be computed exploiting the diagonal state $\nu$
of Eq.\ (\ref{thermas}).
One obtains
\be {\rm
tr}\,\rho^{p}=\prod_{i=1}^{n}\frac{2^p}{f_{p}(\nu^{\downarrow}_i)}
=\frac{2^{pn}}{F_{p}(\gamma )}\; , \label{pgau} \ee
where
\begin{equation}
f_{p}(x)=(x+1)^p-(x-1)^p
\end{equation}
and we have defined $F_p(\gamma) =\prod_{i=1}^{n}f_p(\nu_i)$, in
terms of the symplectic eigenvalues of the covariance matrix
$\gamma$ of $\rho$. A first consequence of \eq{pgau} is that
the purity $\mu$ of a Gaussian state is fully determined by the
symplectic invariant $\det\gamma$ alone: \be
\mu(\rho)=\frac{1}{\prod_{i=1}^n \nu^{\downarrow}_{i}}=\frac{1}{\sqrt{
\det[\gamma] }} \, . \label{purgau} \ee Every R\'enyi entropy, or
$p$-norm respectively, yields an order within the set of density
operators with respect to the purity of the states. Yet a stronger
condition for one state being more ordered than another one is
given by the majorization relation which gives rise to a
half-ordering in state space. A density operator $\rho$ is said to
majorize $\tilde \rho$, i.e., $\rho\succ\tilde\rho$ if
\be\label{Eqmajo}
\sum_{i=1}^r \lambda_i^{\downarrow} \geq
\sum_{i=1}^r \tilde\lambda_i^{\downarrow}
,\ee
for all $r\geq 1$,
where $\lambda_i^{\downarrow}$ is the decreasingly ordered
spectrum of $\rho$. Majorization is the strongest ordering
relation in the sense that if $\rho\succ\tilde\rho$, then $\tr
f(\rho)\leq\tr f(\tilde\rho)$ holds for any concave function $f$
and in particular for every R\'enyi entropy \cite{bathia}. It has
recently been conjectured \cite{lloyd} for a special class of
Gaussian single-mode channels that the maximal output purity is
not only achieved for a Gaussian input state but that the optimal
Gaussian output even majorizes any other possible output state.

\section{Multiplicativity of the maximal output purities}\label{mop}

We define now  for $p\in(1,\infty)$
the Gaussian maximal output $p$-purity of a Gaussian channel
$\Phi_{X,Y}:{\cal G}\rightarrow {\cal G}$,
as
\begin{equation}
\xi_{p}(\Phi_{X,Y})=\sup_{\rho\in \G}\, \|\Phi_{X,Y}(\rho)\|_p \;
,\label{max}
\end{equation}
where the $\sup$ is taken over the set of Gaussian states.
In terms of covariance matrices and of the function $F_p$, one has
\begin{equation}
    \left(\frac{2^{n}}{\xi_{p}(\Phi_{X,Y})}\right)^p=
    \inf_{\rho}
    F_p\left(\phi_{X,Y}(\gamma )\right) \; . \label{maxcov}
\end{equation}
Thus, on the level of second moments, the multiplicativity of the Gaussian
maximal output
$p$-purity under Gaussian inputs
corresponds to the multiplicativity of the infimum of $F_p$
\begin{equation}
\inf_{\gamma }F_p(\phi_{X,Y}(\gamma))=
\prod_{i=1}^n\inf_{\gamma }F_p(\phi_{X_i,Y_i}(\gamma)) \, , \label{infi}
\end{equation}
where the infimum is taken over all covariance matrices.

For finite dimensional systems and the usual definition of maximal
output purity (allowing for input on the whole convex set of trace
class operators), the convexity of the $p$-norms guarantees that
the $\sup$ of \eq{max} can be approached restricting to pure
states. The set $\G$ of Gaussian states is not convex. However,
every Gaussian state still has a convex decomposition into pure
Gaussian states such that it is again sufficient to consider pure
input states only:
\smallskip

\noindent{\bf Lemma 1. --} {\em For any Gaussian channel
$\Phi_{X,Y} :{\cal G}\rightarrow {\cal G}$ and any
$p\in(1,\infty)$, one has
\begin{equation}
\sup_{\rho \in \G}\, \|\Phi_{X,Y}(\rho )\|_p=\sup_{\rho
\in \tilde{\G}}\, \|\Phi_{X,Y}(\rho )\|_p \; . \label{prop1}
\end{equation}
}
\medskip

\noindent{\em Proof.} Consider the Williamson standard form of the
covariance matrix $\gamma=S^T\nu S$. By rewriting this as
\begin{equation}
\gamma= S^T S + S^T(\nu-\ii)S=:\gamma_p+V
\end{equation} one infers from Eq.\
(\ref{averagedgamma}) that the state corresponding to $\gamma$ can
be generated by randomly displacing a pure state with covariance matrix
$\gamma_p$ in
phase space according to a classical Gaussian probability
distribution with covariance $V$. Hence, the state has a convex
decomposition into pure Gaussian states and the Lemma follows from
the convexity of the $p$-norms.$\Box$\smallskip

Let us now consider a channel $\Phi_{X,Y}$ resulting from the tensor
product of the channels $\Phi_{X_i,Y_i}$,
$i=1,...,n$,
\begin{equation}
\Phi_{X,Y}=\bigotimes_{i=1}^n\Phi_{X_i,Y_i},
\end{equation}
acting on Gaussian states associated with
 the tensor product Hilbert space.
Since tensor products in Hilbert spaces correspond to direct sums in phase
space, we have that
$\Phi_{X,Y}=
\Phi_{\oplus X_i,\oplus Y_i}$.
We will say that the Gaussian maximal output $p$-purity of the
channel $\Phi_{X,Y}$ is \emph{multiplicative} if
\begin{equation}
\xi_p(\Phi_{X,Y})=\prod_{i=1}^n
\xi_p(\Phi_{X_i,Y_i})\; .
\end{equation}
Let us remark that this is equivalent to stating that the maximal
output purity can be attained by means of uncorrelated input
states. More precisely, for tensor products of channels one has
that, denoting by ${\cal S}$ the set of product Gaussian states
with respect to each of the modes, the multiplicativity of the
maximal output $p$-purity of the channel $\Phi_{X,Y}$ is
equivalent to
\begin{equation}
\xi_p(\Phi_{X,Y})=\inf_{\rho\in{\cal S}}\|\Phi_{X,Y}(\rho)\|_p \, .
\label{sepa}
\end{equation}

\section{Multiplicativity statements}\label{theo}

From now on, we will mainly restrict to tensor products of
single-mode channels, for which the matrices describing the
channels are direct sums of $2\times 2$ matrices $X_i$ and
$Y_{i}$, $i=1,...,n$: $X=\oplus_{i=1}^n X_i$ and $Y=\oplus_{i=1}^n
Y_i$. Moreover, we will assume that the determinants of the $X_i$
have equal sign. In this case the invariance of the $p$-norms and
of the correlations of quantum states under local unitary
operations can be exploited to simplify the problem, according to
the following.\smallskip

\noindent{\bf Lemma 2. --} {\em Let the determinants of the
matrices $\{X_i:1,...,n\}$ have equal signs. Then the Gaussian maximal
output $p$-purity of the tensor product of single-mode channels
$\Phi_{\oplus X_i,\oplus Y_i}$ is multiplicative if and only if
the Gaussian maximal output $p$-purity of the channel
$\Phi_{\tilde{X},\tilde{Y}}$ is multiplicative, with
\begin{eqnarray}
\tilde{X}&=&\bigoplus_{i=1}^n\tilde{X}_i \, , \label{ides}  \quad
\tilde{Y}=\bigoplus_{i=1}^n\tilde{Y}_i \, , \\
\tilde{X}_i &=&\sqrt{|\det [X_i] | } {\id}_2 \, , \quad
\tilde{Y}_i=\sqrt{\det [Y_i] } {\id}_2. \label{ide}
\end{eqnarray}

}

\noindent{\em Proof.} Let us first reduce the case of negative
determinants to that of $\det [X_i] >0$. To this end we write
$X_i=\sigma_z X^+_i$, so that $\det [X^+_i] =-\det [ X_i ]$. Since
$\theta\gamma\theta$ with $\theta=\bigoplus_{i=1}^n\sigma_z$ is
again an admissible covariance matrix (corresponding to the time
reversed state) and in addition $\theta^2=\ii$, we have indeed that
\be \inf_{\gamma }F_p(X^T \gamma X+Y)
    =  \inf_{\gamma }\, F_p( X^{+T}  \gamma   X^+ + Y)\,
    .
\ee Now let us see how the case $\det [X_i ]>0$ can be reduced to
the standard form given in the Lemma. Due to the unitary
invariance of the $p$-norm we can replace $X,Y$ by $\tilde X=
S'XS$, $\tilde Y=S^T Y S$, with $S,S'$ being any symplectic
transformations and

 \be \label{EqLem2}\inf_{\gamma }F_p(X^T \gamma X+Y)
    =  \inf_{\gamma }\, F_p(\tilde X^{T}  \gamma  \tilde X +\tilde Y)\,
    .
\ee In particular we may choose $S=\bigoplus_{i=1}^n S_i O_i$,
such that $O_i\in SO(2)$  and $S_i \in Sp(2,{\mathbb R})$
bring $Y$ in standard form: $S_iYS_i^T=\sqrt{\det Y_i}\ii= \tilde Y_i$.
Furthermore, we choose $S'$ to consist of blocks
$S_i'=Z_iO_i'$, $O_i\in SO(2)$, $Z_i\in Z(2)$ such that the
orthogonal matrices $O_i, O_i'$ diagonalize $X_i S_i$ in
\begin{equation}
\tilde
X_i=Z_i\left[O_i'(X_i S_i)O_i\right]
\end{equation}
and the squeezing
transformation $Z_i$ gives rise to equal diagonal entries yielding
$\tilde X_i=\sqrt{|\det [X_i] |}\ii$.
According to the block structure of the involved transformations
(corresponding to the direct sum of `local' single mode operations) one has
\be
\prod_{i=1}^n \inf_{\gamma }F_p(X_i^T \gamma X_i+Y_i)
    =  \prod_{i=1}^n
    \inf_{\gamma }F_p(\tilde X_i^T \gamma \tilde X_i+\tilde Y_i)\, .
    \label{eqlem2bis}
\ee
Eqs.~(\ref{EqLem2}) and (\ref{eqlem2bis}) straightforwardly imply that
$\inf_{\gamma }F_p(X^T \gamma X+Y)=
    \prod_{i=1}^n \inf_{\gamma }F_p(X_i^T \gamma
    X_i+Y_i)\label{prod}
$
if and only if
$   \inf_{\gamma }F_p(\tilde X^T \gamma \tilde X+\tilde Y)=
    \prod_{i=1}^n
    \inf_{\gamma }F_p(\tilde X_i^T \gamma \tilde X_i+\tilde Y_i).
$
This
proves the claimed equivalence of the multiplicativity statements.
$\Box$\smallskip

One remark is in order concerning the case $\det X_i=0$ in the
above Lemma. In fact, in this case multiplicativity is trivial and
the maximal $p$-norm output purity does not at all depend on the
$X_i$. To see this note that for two positive matrices $A\geq
B\geq 0$ we have $\nu^{\downarrow}_i(A)\geq \nu^{\downarrow}_i(B)$
\cite{gezajens}, which implies that  $\inf_{\gamma} F_p(X\gamma
X^T+Y)\geq F_p(Y)$. This becomes, however, an equality in the case
$\det X_i=0$ since we can always choose the input state to be a
product of squeezed states such that in the limit of infinite
squeezing $X\gamma X^T\rightarrow 0$.

A first relevant consequence of Lemma 2 follows.\smallskip

\noindent{\bf Proposition 1. --}
 {\em The Gaussian maximal output $p$-purity of a tensor product of $n$
identical single-mode Gaussian
 channels $\Phi_{\oplus_i X,\oplus_i Y}$ is multiplicative for any
$p\in(1,\infty)$. Moreover, the output corresponding to the
optimal product input majorizes any other Gaussian output state of
the channel.}\smallskip

\noindent{\em Proof. } We recall that, because of
Euler decomposition,
the covariance matrix of any pure Gaussian state $\rho$
can be written as $\gamma=O^TZO$, where $O\in
K(n)= Sp({2n,\R})\cap SO(2n)$
is an orthogonal symplectic transformation and $Z\in Z(n)$
corresponds to a tensor product of local squeezings,
$Z={\rm diag}\,(z_1,1/z_1,\ldots ,z_n,1/z_n)$.
Clearly, if $O={\id}$ then the state is uncorrelated.
For a tensor product of identical channels, \eq{ide} holds globally,
$\tilde{X}=x{\id}_{2n}$, $\tilde{Y}=y{\id}_{2n}$.
Therefore, exploiting Lemma 2 and the invariance of $F_p$ under
symplectic transformations, one has
\begin{eqnarray}
\inf_{\gamma}F_p(\tilde{X}\gamma\tilde{X}+\tilde{Y})&\hspace*{-.2cm}=&
\hspace*{-.2cm}\inf_{O\in K(n) \atop Z\in Z(n)} F_p(x^2O^T Z
O+y{\id})\\ \label{EqProp1}&\hspace*{-.2cm}=&\hspace*{-.1cm}
\inf_{Z\in Z(n) } F_p(x^2 Z+y {\id}).
\end{eqnarray}
Due to the block structure of elements in $Z(n)$ this proves the
first part of the proposition.

For the majorization part we exploit the fact that a componentwise
inequality for the symplectic eigenvalues $\nu_i^{\downarrow}\leq
\tilde\nu_i^{\downarrow}$ for all $i$ implies majorization on the
level of density operators, {\em i.e.}~$\rho\succ\tilde\rho$
\cite{gezajens}. The symplectic eigenvalue $\nu_i^{\downarrow}$ of
the output covariance matrix $\gamma'=X^T \gamma X +Y$ is given by
the square root of  the ordinary eigenvalue
$\lambda_i^{\downarrow}$ of the matrix
$\sigma\gamma'\sigma^T\gamma'$ (when appropriately taking degeneracies into
account). Continuing with the expression in
Eq.\ (\ref{EqProp1}) we have thus to consider the dependence of
\begin{eqnarray}
\lambda_i^{\downarrow} ( \sigma\gamma'\sigma^T\gamma' )
&=& x^4+y^2+x^2 y\;\lambda_i^{\downarrow} ( Z+Z^{-1} )\label{eigen}
\end{eqnarray}on $Z$. However, choosing $Z=\ii$ in Eq.~(\ref{eigen})
minimizes all the
eigenvalues simultaneously and thus proves the
desired inequalities between the optimal and any other Gaussian
output state. $\Box$\smallskip

In the following we will investigate the multiplicativity issue in
the case of tensor products of {\it different} Gaussian channels.
To proceed in this direction, we aim to turn our optimization
problem over the non-convex set of Gaussian states into an
analytical one. To do so two simple remarks, giving rise to
alternative parametrizations of pure covariance matrices, will be exploited.
Firstly, because of the Euler decomposition given by \eq{euler2},
the set of the pure $n$-mode covariance matrices can be parametrized by means of
the functions
\begin{equation}
\tilde{\gamma}: ({\mathbbm R^+})^{n}
\times {\mathbbm R }^{n^2}
\longrightarrow
\tilde{\G}
\end{equation}
defined as
\begin{equation}
\tilde{\gamma}(l,z)=\,{e}^{-\sum_{i=1}^{n^2}l_i L_i^T} D(z)
\,{e}^{-\sum_{i=1}^{n^2}l_i L_i}\, , \label{eulpar}
\end{equation}
where the $L_i$, $i=1,...,n^2$,
are the generators of the compact subgroup
$K(n)$ (see App.~\ref{genius})
and $D(z)={\rm diag}\,(z_1,1/z_1,\ldots,z_n,1/z_n)\in Z(n)$,
with $z_i>0 $ for all $i$.
Here, $l=(l_1,...,l_{n^2})$ is a vector of
$n^2$ real parameters while $z=(z_1,...,z_n)$ is a vector
of $n$ real strictly positive parameters.\\
Otherwise,
the set of pure covariance matrices admits the following
parametrization
\be
\hat{\gamma}_{\gamma} (k) =  \,{e}^{\sum_{i=1}^{d}k_i K_i^T} \gamma
\,{e}^{\sum_{i=1}^{d}k_i K_i}\, ,
\ee
where $\gamma$ is an arbitrary pure covariance matrix,
the $K_i$ are the $d=2n^2+n$ generators of the symplectic group
(detailed in App.~\ref{genius}) and $k=(k_1,\ldots,k_d)$ is a real vector
of dimension $d$. We are now in a position to prove our main result.\smallskip

\noindent{\bf Proposition 2. --} {\em The Gaussian maximal output
$p$-purity of a tensor product of single-mode Gaussian channels
$\Phi_{\oplus X_i,\oplus Y_i}$ with $Y_i>0$ and identical $\,\det
[X_i]$\ for all $i$   is multiplicative for any
$p\in(1,\infty)$.}\smallskip

%We will focus on the function $G_p$ defined
%by \eq{defg} with
%$\tilde{X}=\id$ and determined by the symplectic eigenvalues $\{\nu^{\downarrow}_1,...,
%\nu^{\downarrow}_n\}$
%of the output CM $\tilde{\gamma}+\tilde{Y}$.
%
%Change order. First: critical points.
%The first step is to see that all critical points of the
%differentiable function $G_p$
%correspond to local minima.
%
\noindent{\em Proof.} Because of Lemma 2, this multiplicativity
issue is equivalent to the one for the `simplified channel'
$\Phi_{\tilde{X},\tilde{Y}}$, with $\tilde{X}=x\id_{2n}$,
$x=\sqrt{|\det [X_i ] |}$ and $\tilde{Y}=\oplus_i \sqrt{\det [Y_i]}\id_2$
according to Eqs.\ (\ref{ides}, \ref{ide}). For ease of notation,
and since the subsequent argumentation does not depend on the
value of $x$, we will state the proof for $x=1$.

In a first step we aim to show that the infimum of \eq{infi} is
indeed a minimum, that is, the infimum of $F_p
(\Phi_{\ii,\tilde{Y}}(\gamma))$ is achieved for a
defined input and not asymptotically approached in
the non compact set of pure covariance matrices. Therefore, we
will analyse the asymptotic behaviour of the output purity of the
channel $\Phi_{\id,\tilde{Y}}$ in the limiting case of infinite
squeezing. For a given channel of this kind, let us define the
function $G_{p, \tilde{Y}}: ({\mathbbm R^+})^{n}\times {\mathbbm R
}^{n^2} \rightarrow {\mathbbm R}$ as
\begin{equation}
G_{p, \tilde{Y}}(l,z)=( F_p\circ \Phi_{\id, \tilde{Y}}
)(\tilde{\gamma}(l,z)) ,\quad p>1 \, , \label{defg}
\end{equation}
where the function $\tilde{\gamma}(l,z)$ has been defined in
\eq{eulpar}. To show that the function is indeed attained by a
(possibly not unique) given covariance matrix, we address the asymptotic
behaviour of $G_{p,\tilde{Y}}$, showing that its infimum cannot be
asymptotically approached. To see this, let us investigate the
product of the symplectic eigenvalues
\begin{equation}
\prod_{i=1}^{n}(\nu_i^\downarrow)^2=
\det [\tilde{\gamma}(l,z)+ \tilde{Y} ] \; .
\end{equation}
Such a function is periodic in $l$, because these variables are
related to the compact subgroup of $Sp({2n,\R})$ (consisting of
rotations). Thus the domain of the $l\in{\R}^{n^2}$ can be chosen
compact. Therefore, only the cases $z_i\rightarrow 0$ and
$z_i\rightarrow\infty$ have to be considered, for $i=1,...,n$. Let
us take the ordered list of $y_1^\downarrow, ..., y_n^\downarrow$,
where $\tilde{Y}=\oplus_{i=1}^n \tilde{Y}_i$ and $\tilde{Y}_i =
y_i^\downarrow \id_2$. Then, clearly, by the positivity of
$\tilde{Y}$ (implied by the positivity of $Y$),
$y_n^\downarrow>0$. Now, it can be easily shown \cite{bathia} that
for two positive $m\times m$ matrices $A$ and $B$ with ordered
lists of eigenvalues $a_1^\downarrow,...,a_m^\downarrow$ and
$b_1^\downarrow,...,b_m^\downarrow$ respectively, one has
\begin{equation}
    \det[A+B]\ge \prod_{i=1}^m a^{\downarrow}_i b^{\downarrow}_m.
\end{equation}
Therefore,
noticing that the
values
${z_i,1/z_i}$ constitute the spectrum of
$\tilde{\gamma}(l,z)$, we get
\begin{equation}
 \det [\tilde {\gamma}({l},{z}) + \tilde{Y}] \ge
\prod_{i=1}^n (z_i+
 {y}_n^\downarrow)(\frac{1}{z_i}+{y}^\downarrow_n) \, .
\end{equation}
From this it immediately follows that, for any $i=1,...,n$,
\begin{equation}
\lim_{z_i\rightarrow 0}\det[\tilde{\gamma}({l},{z})+\tilde{Y}] =
\lim_{z_i\rightarrow+\infty}
\det[\tilde{\gamma}({l},{z})+\tilde{Y}]= +\infty \, .
\label{detasym}
\end{equation}
\eq{detasym} shows that the product of the sympletic eigenvalues
of $\tilde{\gamma}({l},{z})+\tilde{Y}$ diverges for
$z_i\rightarrow 0$ and $z_i\rightarrow \infty$. Moreover, all
symplectic eigenvalues of $\tilde{\gamma}({l},{z})+\tilde{Y}$ are
clearly positive. This shows that the function $G_{p,\tilde{Y}}$
diverges for $z_i\rightarrow 0$ and $z_i\rightarrow \infty$ for
all $i=1,...,n$. In turn, this means that the global infimum of
the function $F_p\circ \Phi_{\id,\tilde{Y}}(\gamma)$ over the set
of pure covariance matrices is indeed a minimum.

In what follows we show that such a minimum is achieved for $\gamma=\id$, that is
for a manifestely uncorrelated input, thus completing the proof.
Let us first notice that, since $F_p$ is a function of the symplectic eigenvalues,
\begin{equation}
F_p(V^T \gamma V) = F_p(\gamma) \label{inva}
\end{equation}
for all $\gamma\in \tilde{\cal G}$, $V\in Sp(2n,R)$, and all channels
$\Phi_{\id,\tilde{Y}}$.

Now, let us call $M$ the infimum of the function $F_p\circ
\Phi_{\id,\tilde{Y}}$ on the whole space of covariance matrices. Suppose that
$\bar{\gamma}$ is one of the optimal matrices, granting the
minimum of $F_p \circ \Phi_{\id,\tilde{Y}}$. Let us denote by $V$
a symplectic transformation bringing $\bar{\gamma}+\tilde{Y}$ in
Williamson form, so that $V^T\bar{\gamma} V+ V^T \tilde{Y}
V=\oplus \nu^{\downarrow}_i \id_2$. Exploiting \eq{inva}, we have
\begin{equation}
M=F_p(\bar{\gamma}+\tilde{Y})=F_p(V^T \bar{\gamma} V + V^T
\tilde{Y} V) \, . \label{minimo}
\end{equation}
For the sake of simplicity, let us define $\gamma'= V^T \bar
\gamma V$ and $Y'=V^T \tilde{Y} V$. We may write the matrix
$\gamma'$ in terms of $2\times 2$ submatrices as
\begin{equation}
 {\gamma}' =\left(\begin{array}{cccc}
\alpha_1&\beta_{12\;}&\ldots&\beta_{1n}\\
&&&\\
\beta_{12\;}^T&\ddots&\ddots&\vdots\\
&&&\\
\vdots&\ddots&\ddots&\beta_{n-1n}\\
&&&\\
\beta_{1n}^T&\ldots&\beta_{n-1n}^T&\alpha_n
\end{array}\right) \, .
\end{equation}
For a given channel and covariance matrix $\bar{\gamma}$,
let us define the function $H_{p}=
F_p\circ \Phi_{\id,{Y}'}(\hat{\gamma}_{\gamma'}(k)):
\R^{d}\longrightarrow
\R$ (with $d=2n^2+n$)

\begin{equation}
H_{p}(k)=F_p({\rm e}^{\sum_i K_i k_i}V^T \bar{\gamma} V {\rm
e}^{\sum_i K^T_i k_i} + V^T \tilde{Y} V)  \, ,
\end{equation}
with $p\in(1,\infty)$.
Such a function is well defined for any channel and input covariance
matrix. The form of
the $K_i$ is the one given in App.~\ref{genius}.
By definition, $M$ is the minimum of the function $H_{p}(k)$ and,
because of \eq{minimo}, $H_p(0)=M$.
Furthermore, such a function is differentiable in $k=0$.
Therefore,
if the covariance matrix $\bar{\gamma}$ is indeed optimal the function
$H_p(k)$ has to be stationary (critical) in $k=0$.
This constraint is explicitly expressed by
\be
\left.\frac{\partial}{\partial k_i}
    \right|_{k=0}
    F_p({e}^{\sum_{i=1}^{d}k_i K_i^T} {\gamma}'
    {e}^{\sum_{i=1}^{d}k_i K_i}+
    Y')=0 \, , \label{conditio}
\ee
for all $i=1,\ldots,d$. We have that
\begin{eqnarray}
    &&\left. \frac{\partial}{\partial k_i}
    \right|_{k=0}
    ({e}^{\sum_{i=1}^{d}k_i K_i^T} {\gamma}'
    {e}^{\sum_{i=1}^{d}k_i K_i}+ Y' )\nonumber\\
    &&=
%{\gamma}' +{Y}'
    (K_i^T {\gamma'}+ {\gamma'} K_i)
%+O(k_i^2)
\; . \label{firper}
\end{eqnarray}
%
%\begin{equation}
%S^T \bar{\gamma}S+\tilde{Y} = \bar{\gamma}+ {Y}- k_i
%(K_i^T\bar{\gamma}+\bar{\gamma}K_i)+O(k_i^2) \; . \label{firper}
%\end{equation}
Since ${\gamma}'+{Y}'$ is in Williamson
form, we can apply the
results of App.~\ref{symper}.
On using
%Inserting the first order perturbation of
\eq{firper}
%in
and \eq{symvar}
we obtain for the first derivative
of the symplectic eigenvalues
of the output
%
%the variations $\{\delta\nu_j^\downarrow \}$
%eigenvalues
%$\{\nu_j^\downarrow\}$ at first order in $k_i$
\begin{equation}
\left.\frac{\partial}{\partial k_i}
\right|_{k=0}
\nu_j^\downarrow=
 \tr_j{[K_i^T {\gamma'}+ {\gamma'} K_i]}\; ,
\end{equation}
where
$\tr_j$ denotes the trace of the
leading principle submatrix
corresponding to the mode $j$ (see App.~\ref{symper}).
For the derivative of the output
$p$-purity we hence obtain
\begin{equation}
\begin{split}
&\hspace*{-1cm}\left.\frac{\partial}{\partial k_i}
    \right|_{k=0}
    F_p({e}^{\sum_{i=1}^{d}k_i K_i^T} {\gamma}'
    {e}^{\sum_{i=1}^{d}k_i K_i}+
    Y')\\
\hspace*{.8cm}= &F_p({\gamma}'+{Y}') \label{varia}
\sum_{j=1}^{n}
\frac{f'(\nu_j^\downarrow)}{f(\nu_j^\downarrow)}
\,\tr_j{[K_i^T {\gamma}'+ {\gamma}' K_i] } \, .
\end{split}
\end{equation}
where $f'=df/dx$.
In order for ${\gamma}'$ to be the an input corresponding to
a critical point,
%granting the minimal
%$F_p$
%(and the maximal output $p$-purity),
%such variation $\delta F_p$ has to be
this derivative has to be zero for
all
generators $K_i$
of $Sp({2n,\R})$.
%
%null for any
%$K_i$ belonging to the set of generators of $Sp({2n,\R})$. Otherwise, by
%properly choosing
%the sign of $k_i$ in \eq{varia}, it would always be possible to achieve a
%greater purity
%by means of an `infinitesimally close', but different, input CM.
%This is always the case for the antisymmetric $K_i$'s (generating the
%compact subgroup
%of $Sp({2n,\R})$), since the perturbation to the output is in that case
%a commutator (and therefore traceless). On the other hand,
As can be promptly verified exploiting the form
of the generators and
\eq{varia}, the condition of Eq.~(\ref{conditio})
results, for the symmetric $K_i$, in the following
constraints on the submatrices of ${\gamma}'$
$\alpha_i$, and $\beta_{kl}$
\begin{equation}
\alpha_{i}=a_1 {\id}_{2} \, , \quad
\beta_{kl}=\left(\begin{array}{cc}
b'_{kl}&b''_{kl}\\
-b''_{kl}&b'_{kl}
\end{array}\right) \; , \label{condvar}
\end{equation}
for some real $a_i\ge 1$ and $b'_{kl}$, $b''_{kl}\in{\mathbb R}$.
Since ${\gamma}'$ is the covariance matrix of a
pure state,
all its symplectic eigenvalues are equal to $1$, implying that
$|i\sigma{\gamma}'|={\id}_{2n}$. Therefore
\begin{equation}
-\sigma{\gamma}'\sigma{\gamma}'={\id}_{2n} \, .
\label{condpur}
\end{equation}
Applying the previous condition to the submatrices of \eq{condvar} yields
\begin{equation}
a_i^2+\sum_{l\neq i}(b_{il}^{\prime2}+b_{il}^{\prime\prime2})=1
\end{equation}
which is equivalent to
\begin{equation}
a_i=1\; ,\, b'_{kl}=b''_{kl}=0 \, . \label{condfin}
\end{equation}
\eq{condfin} shows that the unique
${\gamma}'$
%of the function
%$F_{p}(\bar{\gamma}+\tilde{Y})$
that is consistent with a critical point
is the identity $\id_{2n}$, corresponding
to the $n$-fold tensor product of a coherent state. It is easy to verify that
the identity also satisfy \eq{varia} for the antisymmetric $K_i$.\\
% in the basis in which $\bar{\gamma}+\tilde{Y}$
%is in Williamson form.
%

Summarizing, we have shown that the unique optimal input
$\bar{\gamma}$ corresponds to $\gamma'=V^T \bar{\gamma}V=\id$,
where $V$ is a symplectic transformation for which 
\begin{equation}
V^T
\bar{\gamma} V+V^T \tilde{Y} V 
\end{equation}
is in Williamson standard form.
However, as the identity and $\tilde{Y}$ themselves are in
Williamson form, it is immediate to see that $V=\id$, yielding
$\bar{\gamma}=\id$, which completes the proof. $\Box$ \smallskip

\noindent For $p=2$ the above multiplicativity result can easily
be extended to products of arbitrary channels acting on any number
of modes without imposing additional constraints on the
determinants of the $X_i$ (apart from being non-zero):\smallskip

\noindent{\bf Proposition 3. --} {\em The maximal Gaussian output
$2$-purity of a tensor product of arbitrary multi-mode Gaussian
channels $\Phi_{\oplus X_i,\oplus Y_i}$ with $\det [X_i]\neq 0$\
for all $i$ is multiplicative.}\smallskip

\noindent{\em Proof.}
%Since the determinants of the matrices $X_i$
%may have different signs we cannot use Lemma 2 directly. However,
%we can utilize the ideas in the proof of Lemma 2 in order to
%reduce the channel to a standard form where $\tilde Y=\bigoplus_i
%y_i\ii_2$ and $\tilde X=\bigoplus_i x_i E_i$, with
%$E_i\in\{\ii_2,\sigma_z\}$ depending on the sign of $\det X_i$.
Because of \eq{purgau}, the Gaussian multiplicativity issue
reduces for $p=2$ to the multiplicativity of the infimum of $\det
[\phi_{{X},{Y}}(\gamma)]$ over all covariance matrices
corresponding to pure Gaussian states. For a given Gaussian
channel $\Phi_{X,Y}$, making use of the Binet theorem ($\det AB
=\det A\det B$) and defining
\begin{equation}
Y'={X}^{-1}{Y}{X}^{-1},
\end{equation}
one gets
\begin{eqnarray}
{\det [\,{X}^T\tilde{\gamma}({l},{z}){X}+{Y} ]} =\det[{X} ]^2
\det[\,\tilde{\gamma}({l},{z})+{Y}'\,] .\nonumber
\end{eqnarray}
However, $Y'$ can be diagonalized to $\tilde{Y}'$ by a symplectic
block matrix, which in turn does not change the determinant: \be
\tilde{Y}'= S Y S^T\;,\quad S=\bigoplus_i S_i\;.\ee Therefore, the
problem is equivalent to verifying the multiplicativity of the
infimum of
\begin{equation}
\det [\tilde{\gamma}({l},{z})+\tilde{Y}']=
F_{2}(\Phi_{\id ,\tilde{Y}'})/4^n,
\end{equation}
which we know to hold true because of Proposition 2.
$\Box$ \smallskip

Note that Proposition 2 also implies multiplicativity for other
multi-mode Gaussian channels. In particular if $X_i=x_i S_i$ are
proportional to symplectic transformations with $x_i>0$, then
multiplicativity holds for any $\Phi_{\oplus X_i,\oplus Y_i}$
within the entire range $p\in(1,\infty)$.
%We just mention that the previous Theorem includes Theorem 1 as a
%particular
%instance.

\section{Comments and outlook}\label{comme}

We have addressed the multiplicativity of the maximal output
$p$-purities of tensor products of Gaussian channels described by
\eq{channel}. We have proved that, restricting to Gaussian inputs,
the maximal output $p$-purities are multiplicative for any $p\in
(1,\infty)$ for single-mode channels with $X=\oplus_{i=1}^n X_i$,
$Y=\oplus_{i=1}^n Y_i$ if $\det X_i$ is the same for $i=1,...,n$,
and that the ordinary `purity', corresponding to $p=2$, is
multiplicative for any generic choice of multi-mode Gaussian
channels. In particular, the maximal output purity is
multiplicative for identical $n$-fold single-mode channels for all
values of $p\in (1,\infty)$ and in this case the optimal product
output (which is independent of $p$) majorizes any other Gaussian
output state.

The restriction to Gaussian states,    formally expressed by
the definition
of Gaussian maximal output purity of \eq{max}, is here motivated by
essentially three arguments. Firstly, the question is interesting in its
own right: Gaussian states have a prominent role in quantum information
and communication with continuous variables, where many
%if not most
protocols completely rely on such states. Colloquially, one may
say that the results indicate that entangled Gaussian input states
suffer more decoherence than uncorrelated ones. The results and
arguments presented in this paper constitute a strong hint towards
the multiplicativity of the maximal output purities of general
products of Gaussian channels under Gaussian inputs. In other
words, it seems plausible that, in a fully Gaussian setting, input
entanglement does not help to better preserve output purity of
quantum channels. Notice also that our proofs of multiplicativity
encompass several instances of interest. In particular,
Proposition 1 represents the case of the subsequent uses of a
single channel, where input correlations could be distributed in
time over the global input. In such an instance our result proves
that Gaussian entangled input states suffer more decoherence than
uncorrelated states. We mention as well that Proposition 3
includes the relevant case of dissipation of multi-mode systems in
Gaussian reservoirs, provided that the coupling to the reservoir
is the same for any mode, but allowing for generally different
reservoir states in different modes [see \eq{bath}].

Secondly, the maximal output purities under Gaussian inputs
readily deliver bounds for the maximal output purities of Gaussian
channels, not restricting to Gaussian inputs. Following the
results presented in Refs.\ \cite{lloyd} one might conjecture that these
bounds are tight and that Gaussian input states are already
optimal.
%The methods presented here could also
%give insights of how to assess the output purities for general inputs.

Thirdly, the issues considered here could be the first steps
towards a general theory of quantum information of Gaussian
states, linking output purities to the Gaussian instance of the
entanglement of formation \cite{GEOF}, and Gaussian versions of
channel capacities. As such, the Gaussian picture would deliver a
convenient and powerful testbed in entanglement theory, in an
instance for which a complete solution for the seemingly unrelated
additivity and multiplicativity problems may be anticipated. It is
the aim for future work to establish this connection in
generality.

\subsection*{Acknowledgements}

This work has been supported by the DFG (Schwerpunktprogramm QIV, SPP
1078)
and the European Commission (IST-2001-38877). We thank K.M.R.\
Audenaert for motivating us to make these notes public, and
M.B.\ Plenio for discussions.

\appendix

\section{Generators of the symplectic group}\label{genius}
As can be easily verified from the expression of the
condition
$S^T\sigma S=\sigma$, the symplectic group $Sp({2n,\R})$ is generated
by those matrices which can be written as $K=\sigma J$,
where $J$ is a symmetric $2n\times 2n$ matrix \cite{prama}.
The antisymmetric generators result in orthogonal symplectic
transformations, giving
rise to the compact subgroup $K(n)=Sp({2n,\R})\cap SO(2n)$. Such a subset of
transformations is constituted by `energy preserving' or passive
operations. In contrast, the symmetric
generators generate the non compact subset of the group (made up of active
transformations, like squeezings).
A basis of such generators can be built by means of transformations
affecting only $1$ or $2$ modes at a time.
We define
\begin{equation}
\beta=\left(\begin{array}{cc}
1& 0\\
0&-1
\end{array}\right) \, ,\quad
\delta=\left(\begin{array}{cc}
0&1\\
1&0
\end{array}\right) \,
\end{equation}
and recall the definition of $\sigma$ in \eq{symform} (to be
understood in the single mode, $2\times 2$ instance).
Single mode transformations
are generated by
\begin{equation}
\sigma \; ,\quad
-\sigma\beta=\delta \; ,\quad
\sigma \delta=\beta \; , \label{1gen}
\end{equation}
where $\sigma$ generates the compact single mode rotations while
$\beta$ and $\delta$ generate single mode squeezings.\\
Two-mode transformations (corresponding to the compact set)
are generated by
\begin{equation}
\left(\begin{array}{cc}
0&\sigma\\
\sigma&0
\end{array}\right) \, , \quad
\left(\begin{array}{cc}
0&-{\id}\\
{\id}&0
\end{array}\right) \; . \label{2gen1}
\end{equation}
Whereas two-mode transformations (corresponding to the non-compact set)
are generated by
\begin{equation}
\left(\begin{array}{cc}
0&\delta\\
\delta&0
\end{array}\right) \, , \quad
\left(\begin{array}{cc}
0&\beta\\
\beta&0
\end{array}\right) \; .\label{2gen4}
\end{equation}
The complete set of generators 
\begin{equation}
\{K_i: i=1,..., 2n^2+n\}
\end{equation}
 is described by
\eq{1gen} for any mode and by Eqs.~(\ref{2gen1}) and (\ref{2gen4}) for
any couple of modes.
The total number of independent generators is
\begin{equation}
3n+4n(n-1)/2=2n^2+n.
\end{equation}
The number of generators of the compact subgroup,
which we refer to as $\{L_i:i=1,...,n^2\}$ in this
paper, is   $n+2n(n-1)/2=n^2$.

\section{Symplectic perturbations}\label{symper}

We consider a covariance matrix of $n$ modes in Williamson form
\begin{equation}
    \gamma =
    \bigoplus_{j=1}^n
    \left(
    \begin{array}{cc}
    \nu_j^\downarrow & 0 \\
    0 & \nu_j^\downarrow
    \end{array}
    \right),
\end{equation}
and investigate
the variations of
the symplectic eigenvalues $\nu_1^\downarrow,...,\nu_n^\downarrow$
under an additive perturbation.
Let us consider
    $\gamma + k P$, with $k\in \R^+$, and $P$ being a symmetric
$2n\times 2n$ matrix, partitioned in terms
of $2\times 2$ submatrices $P_{ij}$ as
\begin{equation}
P =\left(\begin{array}{cccc}
P_{11}&P_{12\;}&\ldots&P_{1n}\\
&&&\\
P_{12\;}^T&\ddots&\ddots&\vdots\\
&&&\\
\vdots&\ddots&\ddots&P_{n-1n}\\
&&&\\
P_{1n}^T&\ldots&P_{n-1n}^T&P_{nn}
\end{array}\right) \; .
\end{equation}

The eigenvalues of the matrix $i\sigma \gamma $ are given by
\begin{equation}
(+\nu_1^\downarrow,-\nu_1^\downarrow,\ldots,+\nu_n^\downarrow,-\nu_n^\downarrow),
\end{equation}
with eigenvectors
\begin{eqnarray}
{v_{j+}}&=&(0,\ldots,0,\underbrace{i,1,}_{{\rm mode}\; j}0,\ldots,0)^T  , \\
{v_{j-}}&=&(0,\ldots,0,\underbrace{1,i,}_{{\rm mode}\; j}0,\ldots,0)^T  ,
\end{eqnarray}
for $j=1,\ldots,n$,
so that $i\sigma \gamma {v_{j\mp}}=
\mp \nu_j^\downarrow  {v_{j\mp}}$.
Now, one has
\begin{eqnarray}
    \left.
    \frac{d}{d k}
    \right|_{k=0}
    \nu_j^\downarrow&=&
    v_{j+}
    (i\sigma P) v_{j+}^T
    =
    v_{j+}
    P v_{j+}^T=
    \tr {P_{jj}}\nonumber\\
    &=&\tr_j {P} \; , \label{symvar}
\end{eqnarray}
where we have defined $\tr_j$ as the trace of the leading
submatrix associated to mode $j$.
The first order derivative
of the symplectic eigenvalue $\nu_j^\downarrow$
is just given by the trace of the
$2\times 2$
principal submatrix related to mode $j$
of the matrix embodying the
perturbation.

%\end{multicols}


\begin{thebibliography}{99}
\bibitem{GeneralRemark}
    The general multiplicativity of the maximal output purity for all values
    of $p\in(1,\infty)$,
    conjectured in Ref.\ \cite{amosov}, has
    been disproved by explicit counterexamples in Ref.\ \cite{holwer}.
    However, such a multiplicativity property has
    been shown to hold for a broad variety of channels \cite{Multiplicative}
    and no counterexamples are at present known for $p<4.778$.
\end{thebibliography}
\end{document}